\documentclass[fleqn,10pt]{wlscirep}
\usepackage[utf8]{inputenc}
\usepackage[T1]{fontenc}
\title{Rail journey cost calculator for Great Britain}

\author[1, 2,*,a]{Federico Botta}
\affil[1]{Department of Computer Science, University of Exeter, Exeter, UK}
\affil[2]{The Alan Turing Institute, London, UK}
\affil[*]{f.botta@exeter.ac.uk}
\affil[a]{\url{https://orcid.org/0000-0002-5681-4535}}


\begin{abstract}
Accessibility of different places, such as hospitals or areas with jobs, is important in understanding transportation systems, urban environments, and potential inequalities in what services and opportunities different people can reach. Often, research in this area is framed around the question of whether people living in an area are able to reach certain destinations within a prespecified time frame. However, the cost of such journeys, and whether they are affordable, is often omitted or not considered to the same level. Here, we present a Python package and an associated data set which allows to analyse the cost of train journeys in Great Britain. We present the original data set we used to construct this, the Python package we developed to analyse it, and the output data set which we generated. We envisage our work to allow researchers, policy makers, and other stakeholders, to investigate questions around the cost of train journeys, any geographical or social inequalities arising from this, and how the transport system could be improved.  
\end{abstract}
\begin{document}

\flushbottom
\maketitle
%
%

\section{Introduction}
Being able to travel to jobs, or services such as hospitals and food stores, is a basic need that we should all be able to achieve \cite{kaufmann2004motility, cass2005social}. However, the discussion often revolves around the accessibility of public transport or private vehicles to reach these destinations in a reasonable time, or indeed, at all \cite{duran2016estimation, liao2020disparities, pereira2019future, wu2021urban}. The affordability of such journeys has often been overlooked, and a comprehensive understanding of the social exclusion resulting from the cost of journeys is still missing \cite{mattioli2018vulnerability, titheridge2014transport, TfN, crisp2018tackling}.
\\\\ A better understanding of the economic aspect of public transport journeys also holds significant importance in the transition towards net zero to reduce the reliance on car journeys \cite{edenhofer2015climate}. Previous research has highlighted how low-income, deprived communities are not being reached by the sustainable transport agenda, and many living in those areas can be considered ``forced car owners'' \cite{curl2018household}.
\\\\ Here, we present a novel Python package, with an associated data set, which provides a starting point to study the cost of train journeys in Great Britain. We envisage the code and data presented here to be useful to all practitioners, academics and policy makers alike, interested in this topic. We also believe that our work here provides further evidence that coupling openly available data with openly available tools can help maximise the usability and impact of analysis, facilitating reproducible research \cite{botta2023packaging}.


\section{Data}
\subsection{ATOC fares data}
Data on rail fares were retrieved in February 2022 from the Rail Delivery Group (RDG), which is the British rail industry membership body bringing together rail companies and passenger groups in Great Britain. The RDG releases data on timetable and fares for all trains in Great Britain to encourage reuse and promote rail travel: \url{data.atoc.org}. Here, we particularly focus on the fares data, but a complete description of both the timetable and fares data sets is available at the URL above. The fares data set contains detailed information for all rail journeys in Great Britain. It is important to highlight that the fares structure of rail tickets in Great Britain is particularly complex, with a wide range of options on how fares can be constructed for journeys. Fares can depend on time of travel, rail company, specific routing of the journey, and whether the ticket has been booked in advance or not; split ticketing, where a passenger travels from station A to station B by purchasing a series of tickets for intermediate stations (e.g. A to C, C to D, ...B), can, and often does, affect the price of the journey and, perhaps somewhat counter-intuitively, can make the overall journey cheaper; finally, due to the number of rail operating companies, there is a relatively large number of discounts and railcards available to passengers, which can make journeys cheaper.

At a high level, the fares data set contains fares for all pairs of points in the rail network in a fare and flow format: each potential journey between a pair of points in the rail network is assigned a flow id, and each flow id is then assigned a fare. Note that we refer to flows as between pairs of points in the rail network rather than between pairs of stations; this is because, to simplify the format in which the data set is released by the RDG, a flow can either be between a specific pair of stations (e.g. London Liverpool Street and Manchester Piccadilly), or between clusters of stations. Clusters of stations are used to group together stations for which the cost of a flow from that cluster to the destination is the same for any station in the starting cluster. This is particularly useful for long journeys, where starting from two stations which are very close to one another (in relationship to the overall distance of the journey) makes no difference in the price of the journey. Note that a flow can be between pairs of stations, from a cluster to an individual station, from an individual station to a cluster, or between pairs of clusters. Each train station can, and typically does, belong to many clusters, depending on the specific flows. This allows clusters to encompass stations in a smaller or larger geographical area depending on the flow (typically long distance journeys are associated with geographically large clusters); it is also worth mentioning that there are some further subtleties to the fares data such that stations can also be grouped together in station groups, which are distinct from station clusters, and which are only used for naming purposes (and station groups can be part of clusters themselves). Each flow id, regardless of whether it is from/to a station or a cluster, can be assigned to multiple fares, which correspond to the different ticket types which are available for that flow; these can be, for instance, one-way any-time single tickets, return tickets, advance, or other special discounted tickets. Thus, to identify the fare for a specific journey between two stations and for a specific ticket type, we have to parse all possible flows in the data, identify the one corresponding to the pair of stations (which, in turn, may require breaking down station clusters and groups down to individual stations), and then retrieve from the fares data the entry corresponding to a specific ticket code and flow id. This will give us the cost of that journey. However, it is important to emphasise that this pipeline may not result in an individual fare, because typically there are multiple flows associated between pairs of stations, particularly for long distance journeys. This can be for a number of reasons, such as the journey taking different routes, or relying on different rail companies for connections. Below, we will mainly focus on the lowest possible fare amongst those available for a flow, but note that this will typically not be the most convenient or efficient way of travelling between stations, particularly for long journeys. 
\subsection{NAPTAN data}
We retrieve data from the National Public Transport Access Nodes (NaPTAN), which is Great Britain's official data set of all public transport access nodes:\\ \url{www.gov.uk/government/publications/national-public-transport-access-node-schema}. In particular, the NaPTAN data set contains the location of all public transport access nodes, such as bus, rail, tram, and other services. Here, we specifically use the NaPTAN data on the location of rail stations in Great Britain.
\subsection{Department for Transport's Journey Time Statistics}
We retrieve data from the UK Department for Transport (DfT) on journey times to key services, such as food stores, health care and employment centres: \url{www.gov.uk/government/collections/journey-time-statistics}. The specific content of this data set is beyond the scope of this paper, but, to summarise, it contains statistics on journey times by different modes of transport to a range of key services. Of interest to us here are those specific key services; their location is released as part of this data set by DfT, and we use this to be able to calculate the cost of accessing such services by rail.

\begin{figure*}
    \centering
    \includegraphics[width = \textwidth]{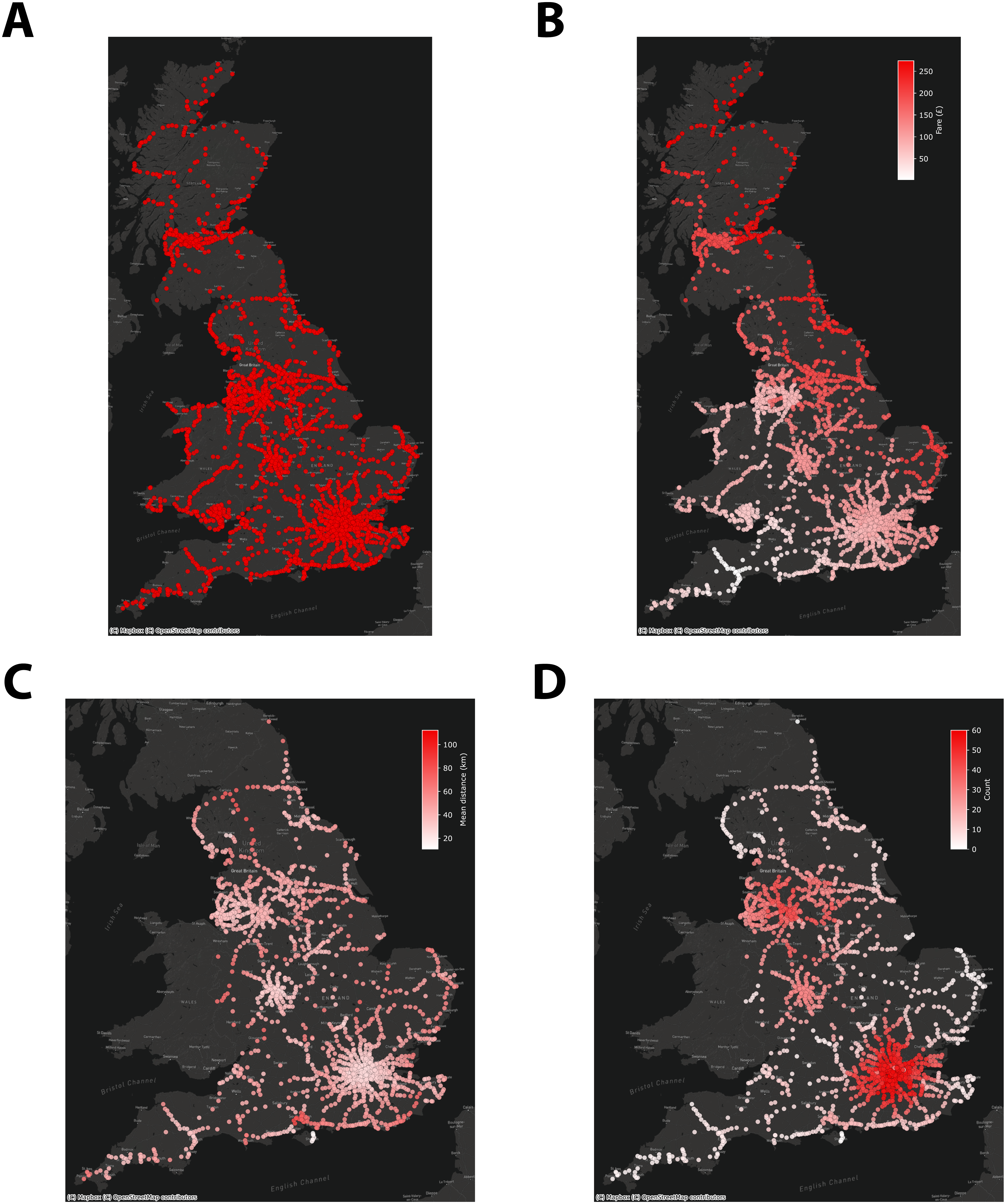}
    \caption{\textbf{Maps $\vert$} We depict an overview of the rail fares data set and the insight that can be gained analysing it. (A) We show the location of all rail stations in Great Britain. (B) We retrieve data on the fares of journeys between pairs of stations. Here, we depict the cost of travelling to any rail station in Great Britain starting from Exeter St David's, the main rail station in Exeter in the South-West of England. (C) We provide an initial exploration of the data on rail fares. Here, we show the mean distance (in kilometres) that you can travel starting from any rail station with a maximum budget of \textsterling{25} with a one-way any-time single fare. Visual inspection suggests a clear distinction between urban and rural areas. (D) The cost of public transport may have an impact on the kind of services and opportunities people may be able to access. Here, we show the number of hospitals (in England) you can reach from a rail station with a maximum budget of \textsterling{25} with a one-way any-time single fare. Note that we consider an hospital to be accessible by train if it is within a 5 km radius from a train station.}
    \label{fig:maps}
\end{figure*}
\begin{figure*}
    \centering
    \includegraphics[width = \textwidth]{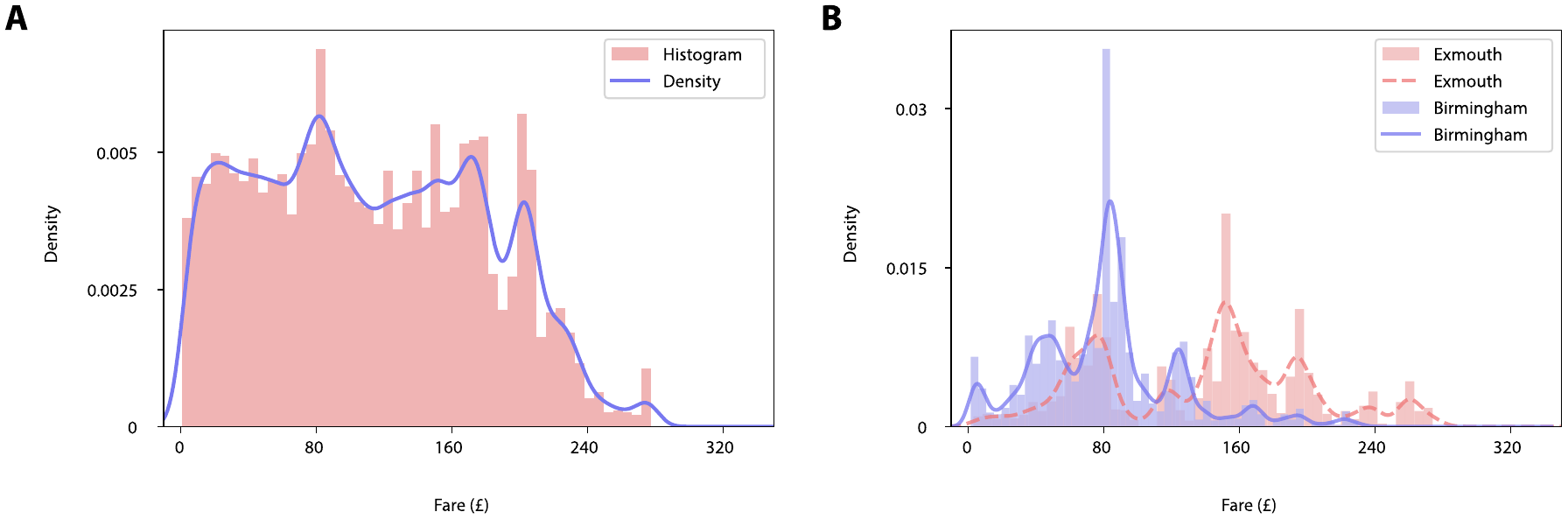}
    \caption{\textbf{Distribution of rail fares $\vert$} We provide an initial investigation into the rail fares data to demonstrate the capability of the Python package described here. (A) We depict the distribution of all rail fares corresponding to the cheapest one-way any-time single fare available for travelling between each pair of stations in Great Britain. (B) The Python package presented here allows for easy comparisons between the rail fares for different stations. Here, we show the distribution of fares for a one-way any-time single ticket leaving from two stations: Exmouth, a small sea-side town in the South-West of England; and Birmingham (Birmingham New Street station), the second largest city in the UK and a major hub in the middle of the country. Visual inspection highlights the differences in the fares.}
    \label{fig:distributions}
\end{figure*}
\section{Code}
\label{sec:code}
The Python implementation of this package has been developed in a Poetry environment running Python 3.9.1, and is available on GitHub \url{https://github.com/FedericoBotta/railfares}. The core of the package is in the \texttt{data\_parsing.py} file, which contains all the key functionalities needed to parse the RDG fares data and generate the output data set described in section \ref{sec:od}, as well as processing the data to generate the figures in this paper. Each function is documented to explain its key functionality. The script \texttt{functionalities.py} contains a few useful functions that are used in other scripts but are not key in processing the data. Finally, \texttt{download\_data.py} allows the user to download all relevant input data made available here.

Alongside the Python package, we also release some Python scripts which use the package to analyse the data:
\begin{itemize}
    \item \texttt{od\_calculation.py} uses the functionalities of the package to generate the data described in section \ref{sec:od}
    \item \texttt{od\_mapping.py} can be used to generate maps such as that of Figure \ref{fig:maps}B
    \item \texttt{number\_hospitals\_multiple\_budgets.py} and \texttt{number\_employment\_centres\_multiple\_budgets.py} can be used to calculate the number of hospitals and employment centres (as specified in DfT's Journey Time Statistics data set) which can be reached with a range of budgets
    \item \texttt{tutorial.py} contains a brief tutorial on how to use the package
\end{itemize}

A brief tutorial script is also provided on the GitHub page to demonstrate basic functionalities and usage of this package (\url{https://github.com/FedericoBotta/railfares}).

\section{Output data}
\label{sec:od}
Using the data sets and Python package described above, we can construct a new data set containing exclusively the minimum fares for travelling between each pair of rail stations in Great Britain using a one-way any-time single ticket. We extract from the RDG data the cost of journeys for individual pairs of stations and release this as a new data set, which can be used to investigate the minimum cost of journeys between pairs of stations. Figure \ref{fig:maps} depicts an overview of this data set, where each circle represents a train station. Panel B shows the cost of travelling to any station in Great Britain starting from Exeter St David's, a city in the South West of England. As we would expect, the cost of travelling increases the further you move from the starting station, with the north of England, Scotland, and parts of the East of England showing the highest prices. Panel C depicts the mean distance you can travel from each station in England with a maximum budget of \textsterling{25}. We can easily see a clear distinction between stations in urban and rural areas. 

The Python package also allows to calculate some metrics about the services and opportunities that can be accessed by train with a range of budgets. The relevant scripts are available on the GitHub repository, and briefly described in section \ref{sec:code} above. Figure \ref{fig:maps}D shows the number of hospitals you can reach from each station with a maximum budget of \textsterling{25}. Note that we assume a hospital can be reached by train if it is within a five kilometres radius of a train station. Visual inspection easily reveals that people in cities experience access to a larger number of hospitals. It is worth highlighting that this is for illustrative purposes of the data and Python package only. However, this example demonstrates how the Python package can easily calculate the reachability, by cost, of destinations of interest as long as the user has data on their location.

\begin{figure}
    \centering
    \includegraphics[width = 0.5\textwidth]{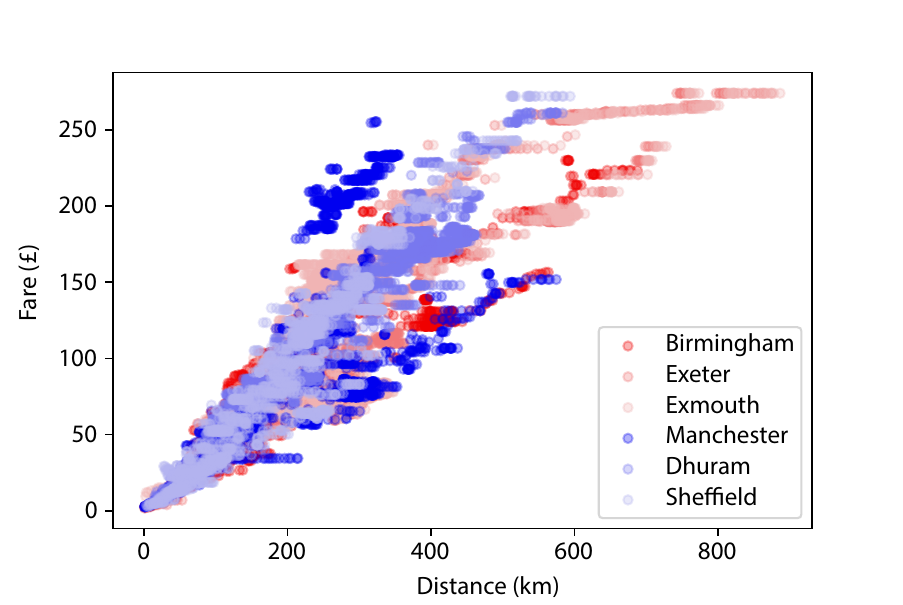}
    \caption{\textbf{Comparison of distance and fares across multiple stations $\vert$} We further analyse the data to provide additional insight into the data set and Python package released here. We highlight the relationship between the distance travelled in a journey and the corresponding cost of the journey for six different stations in England: Birmingham New Street; Exeter St David's; Exmouth; Manchester Piccadilly; Dhuram; Sheffield. We picked these stations for their diversity in terms of location and size of associated city.}
    \label{fig:dist_fare}
\end{figure}

The data on the minimum fares and the data on the number of hospitals, employment centres, and town centres, is made available as a release on the GitHub page of the project (\url{https://github.com/FedericoBotta/railfares/releases/download/v1/Results.zip}), and a description of the files format is given in the README file (\url{https://github.com/FedericoBotta/railfares/blob/main/README.md}).
\begin{table*}
\caption{\textbf{Distribution summary statistics} We provide here the summary statistics of the distribution of fares depicted in Fig.~\ref{fig:distributions}.}
\label{tab:dist_summary}
\centering
\vspace{10pt}
\begin{tabular}{@{}cccccc@{}} \toprule 
Mean & Median & Min & Max & Lower quartile & Upper quartile \\ \midrule
\textsterling{111.64} & \textsterling{106.7} & \textsterling{0.65} & \textsterling{283.3} & \textsterling{55.5} & \textsterling{166.0}\\
\bottomrule
\end{tabular}
\end{table*}
To demonstrate the capability of the Python package, and the associated data set, Figures \ref{fig:distributions} and \ref{fig:dist_fare} provide an overview of some general properties of the data. Fig.~\ref{fig:distributions}A depicts the distribution of (any-time, single) ticket prices for all journeys in Great Britain, and Table \ref{tab:dist_summary} provides the summary statistics for this distribution. To further investigate differences between stations, Fig.~\ref{fig:distributions}B shows the distribution for two stations in quite different locations: Birmingham New Street is the main station in the city of Birmingham, the UK's second largest city, which is located in the middle of England and acts as a transport hub connecting different parts of the country together; Exmouth is a small seaside town in the south-west of England, which is a relatively rural part of the country with significantly fewer train links compared to Birmingham. Finally, Fig.~\ref{fig:dist_fare} depicts the relationship between the cost of tickets and distance travelled for six different train stations in England: Birmingham New Street; Exeter St David's; Exmouth; Manchester Piccadilly; Durham; Sheffield. These stations are in significantly different locations, with Birmingham and Manchester being large cities and transportation hubs; Sheffield, Durham and Exeter are cities in different areas of England and of varying sizes, and Exmouth a small seaside town. Visual inspection reveals a positive correlation between price and distance, which is to be expected, but also some interesting variation across stations as well as some clusters in the points. We anticipate that the work presented here will enable the study of these, and other, interesting features.

\section{Conclusion}
We presented a first effort to enable data science practitioners to study the cost of train journeys across the whole of Great Britain. Our work builds on official data on rail fares and we release a Python package and an associated data set that allows to calculate the cost of journeys between pairs of stations, investigate accessibility of different types of services, and analyse spatial variations in the fares.
\\\\ Undoubtedly, there are a number of limitations and challenges too. The current version of the Python package requires users to download the most recent version of the data from the Rail Delivery Group, should they wish to use the latest fares. The complex ticketing system in Great Britain requires to make some simplified choices in order to make comparisons, but this is likely to result in some inconsistencies at times. However, we believe our Python package allows to consider different ticket types to those discussed here, and the interested researcher should be able to edit the relevant parameters in the calls to the functions calculating the cost of journeys. This is already possible in the current version of the package, albeit it could probably be made more explicit and easier to do. Finally, it is clear that the true cost of a journey is more complex than what discussed here, since it should include the cost of reaching a train station, any additional travel at the destination, as well as further barriers to travelling, such as accessibility of train stations. We envisage the work presented here as a starting point for the interested community to build upon.

\section*{Acknowledgements}
This work is funded by the Economic and Social Research Council (ESRC) \& ADR UK as part of the ESRC-ADR UK No.10 Data Science (10DS) fellowship in collaboration with 10DS and ONS (grant number ES/W003937/1). The views expressed are those of the author and may not reflect the views of 10DS, the Office for National Statistics and the wider UK government.

\bibliography{references.bib}

\begin{thebibliography}{10}
\urlstyle{rm}
\expandafter\ifx\csname url\endcsname\relax
  \def\url#1{\texttt{#1}}\fi
\expandafter\ifx\csname urlprefix\endcsname\relax\def\urlprefix{URL }\fi
\expandafter\ifx\csname doiprefix\endcsname\relax\def\doiprefix{DOI: }\fi
\providecommand{\bibinfo}[2]{#2}
\providecommand{\eprint}[2][]{\url{#2}}

\bibitem{kaufmann2004motility}
\bibinfo{author}{Kaufmann, V.}, \bibinfo{author}{Bergman, M.~M.} \& \bibinfo{author}{Joye, D.}
\newblock \bibinfo{journal}{\bibinfo{title}{Motility: Mobility as capital}}.
\newblock {\emph{\JournalTitle{International journal of urban and regional research}}} \textbf{\bibinfo{volume}{28}}, \bibinfo{pages}{745--756} (\bibinfo{year}{2004}).

\bibitem{cass2005social}
\bibinfo{author}{Cass, N.}, \bibinfo{author}{Shove, E.} \& \bibinfo{author}{Urry, J.}
\newblock \bibinfo{journal}{\bibinfo{title}{Social exclusion, mobility and access}}.
\newblock {\emph{\JournalTitle{The sociological review}}} \textbf{\bibinfo{volume}{53}}, \bibinfo{pages}{539--555} (\bibinfo{year}{2005}).

\bibitem{duran2016estimation}
\bibinfo{author}{Dur{\'a}n-Hormaz{\'a}bal, E.} \& \bibinfo{author}{Tirachini, A.}
\newblock \bibinfo{journal}{\bibinfo{title}{Estimation of travel time variability for cars, buses, metro and door-to-door public transport trips in santiago, chile}}.
\newblock {\emph{\JournalTitle{Research in Transportation Economics}}} \textbf{\bibinfo{volume}{59}}, \bibinfo{pages}{26--39} (\bibinfo{year}{2016}).

\bibitem{liao2020disparities}
\bibinfo{author}{Liao, Y.}, \bibinfo{author}{Gil, J.}, \bibinfo{author}{Pereira, R.~H.}, \bibinfo{author}{Yeh, S.} \& \bibinfo{author}{Verendel, V.}
\newblock \bibinfo{journal}{\bibinfo{title}{Disparities in travel times between car and transit: Spatiotemporal patterns in cities}}.
\newblock {\emph{\JournalTitle{Scientific reports}}} \textbf{\bibinfo{volume}{10}}, \bibinfo{pages}{4056} (\bibinfo{year}{2020}).

\bibitem{pereira2019future}
\bibinfo{author}{Pereira, R.~H.}
\newblock \bibinfo{journal}{\bibinfo{title}{Future accessibility impacts of transport policy scenarios: Equity and sensitivity to travel time thresholds for bus rapid transit expansion in rio de janeiro}}.
\newblock {\emph{\JournalTitle{Journal of Transport Geography}}} \textbf{\bibinfo{volume}{74}}, \bibinfo{pages}{321--332} (\bibinfo{year}{2019}).

\bibitem{wu2021urban}
\bibinfo{author}{Wu, H.} \emph{et~al.}
\newblock \bibinfo{journal}{\bibinfo{title}{Urban access across the globe: an international comparison of different transport modes}}.
\newblock {\emph{\JournalTitle{npj Urban Sustainability}}} \textbf{\bibinfo{volume}{1}}, \bibinfo{pages}{16} (\bibinfo{year}{2021}).

\bibitem{mattioli2018vulnerability}
\bibinfo{author}{Mattioli, G.}, \bibinfo{author}{Wadud, Z.} \& \bibinfo{author}{Lucas, K.}
\newblock \bibinfo{journal}{\bibinfo{title}{Vulnerability to fuel price increases in the uk: A household level analysis}}.
\newblock {\emph{\JournalTitle{Transportation Research Part A: Policy and Practice}}} \textbf{\bibinfo{volume}{113}}, \bibinfo{pages}{227--242} (\bibinfo{year}{2018}).

\bibitem{titheridge2014transport}
\bibinfo{author}{Titheridge, H.}, \bibinfo{author}{Mackett, R.~L.}, \bibinfo{author}{Christie, N.}, \bibinfo{author}{Oviedo~Hern{\'a}ndez, D.} \& \bibinfo{author}{Ye, R.}
\newblock \bibinfo{title}{Transport and poverty: a review of the evidence} (\bibinfo{year}{2014}).

\bibitem{TfN}
\bibinfo{title}{Transport-related social exclusion in the north of england} (\bibinfo{year}{2022}).

\bibitem{crisp2018tackling}
\bibinfo{author}{Crisp, R.} \emph{et~al.}
\newblock \bibinfo{title}{Tackling transport-related barriers to employment in low-income neighbourhoods} (\bibinfo{year}{2018}).

\bibitem{edenhofer2015climate}
\bibinfo{author}{Edenhofer, O.}
\newblock \emph{\bibinfo{title}{Climate change 2014: mitigation of climate change}}, vol.~\bibinfo{volume}{3} (\bibinfo{publisher}{Cambridge University Press}, \bibinfo{year}{2015}).

\bibitem{curl2018household}
\bibinfo{author}{Curl, A.}, \bibinfo{author}{Clark, J.} \& \bibinfo{author}{Kearns, A.}
\newblock \bibinfo{journal}{\bibinfo{title}{Household car adoption and financial distress in deprived urban communities: A case of forced car ownership?}}
\newblock {\emph{\JournalTitle{Transport Policy}}} \textbf{\bibinfo{volume}{65}}, \bibinfo{pages}{61--71} (\bibinfo{year}{2018}).

\bibitem{botta2023packaging}
\bibinfo{author}{Botta, F.}, \bibinfo{author}{Lovelace, R.}, \bibinfo{author}{Gilbert, L.} \& \bibinfo{author}{Turrell, A.}
\newblock \bibinfo{journal}{\bibinfo{title}{Packaging code for reproducible research in the public sector}}.
\newblock {\emph{\JournalTitle{arXiv preprint arXiv:2305.16205}}}  (\bibinfo{year}{2023}).

\end{thebibliography}

\end{document}